\documentclass[%
 aapm,
 mph,%
 amsmath,amssymb,
 reprint,%
]{revtex4-2}

\usepackage{graphicx}
\usepackage{svg}
\usepackage{float}
\usepackage{mathptmx}
\usepackage{dcolumn}
\usepackage{bm}
\usepackage{hyperref}
\usepackage{subfigure}
\usepackage[english]{babel}
\addto{\captionsenglish}{%
  
}

\begin{document}

\preprint{}

\title{Water permeability in nanopores: when size, shape, and charge matter}

\author{Jo\~ao P. K. Abal}
\email{joao.abal@ufrgs.br}
\affiliation{Institute of Physics, Federal University of Rio Grande do Sul, 91501-970, Porto Alegre, Brazil}

\author{Marcia C. Barbosa}
\email{marcia.barbosa@ufrgs.br}
\affiliation{Institute of Physics, Federal University of Rio Grande do Sul, 91501-970, Porto Alegre, Brazi}

\date{\today}

\begin{abstract}
Nanoscale materials are a promising desalination technology. While fast water flow in nanotubes is well understood, this is not the case for water permeability in single-layer membranes. The physical-chemical balance between nanopore size, shape, and charge might be the answer.
\end{abstract}


\maketitle
\section{\label{sec:intro}introduction}

A key challenge in the desalination process is creating a membrane where water is able to flow through and rejects salt ions. The most common process to separate salt from water is reverse osmosis, which consists of employing pressure to filter salt water. This process has two key parts:  overcome the osmotic pressure and produce an efficient selective membrane. While the energy of overcoming the osmotic pressure has little space for improvement, selectivity represents the desalination process frontier. The difficulty is that, in traditional polymeric membranes, a large enhancement in water permeability through increased pore size and permeability also implies an increase in salt permeability, which spoils selectivity. Therefore, a new physical phenomenon to increase water flow for small-sized pores was required~\cite{VOUTCHKOV20182,Alvarez2018,Werber2016}. 

This new phenomenon was the enhancement water flow inside nanoscale materials~\cite{hummer2001,majumder2005,holt2006}. A number of studies have shown that sub-nanometer pores act as a highly selective and permeable filtration membrane with greater efficiency than current state-of-the-art polymer-based filtration membranes. The first system in which this property was observed was in carbon nanotubes (CNT). For diameters under 2nm \cite{majumder2005,holt2006}, water molecules exhibit a flow five orders of magnitude larger than those observed in polymer-made membranes. The physical mechanism behind this enhanced mobility is the smooth inner hydrophobic surface of CNTs, which lubricates and speeds up a near-frictionless water transport~\cite{kalra2002}. The drawback is that hydrophobic carbon surfaces, even though not as frictionless for salt as it is for water, fail to repeal salt. Therefore, only CNTs with sub-0.9nm diameters are able to exhibit acceptable rejection rates~\cite{song2009}. In order to circumvent this problem, charged groups were added to the nanotube. The electrostatic forces added through the hydrophilic groups increased salt rejection, but decreased water velocity as well. Surface roughness produced by the hydrophilic groups also led to a reduction in water mobility~\cite{C8TA10941A}. Also, the hydrophilic substrate changes the behavior of water at least in layers very
close to it~\cite{gallo2017}. Even though the physical mechanism behind the fast flow of water in CNTs is well understood, membranes based on it have been limited by low salt rejection rates and the difficulty of producing highly aligned and high-density CNT arrays.

\begin{figure}[hbt!] \centering
\includegraphics[width=8cm]{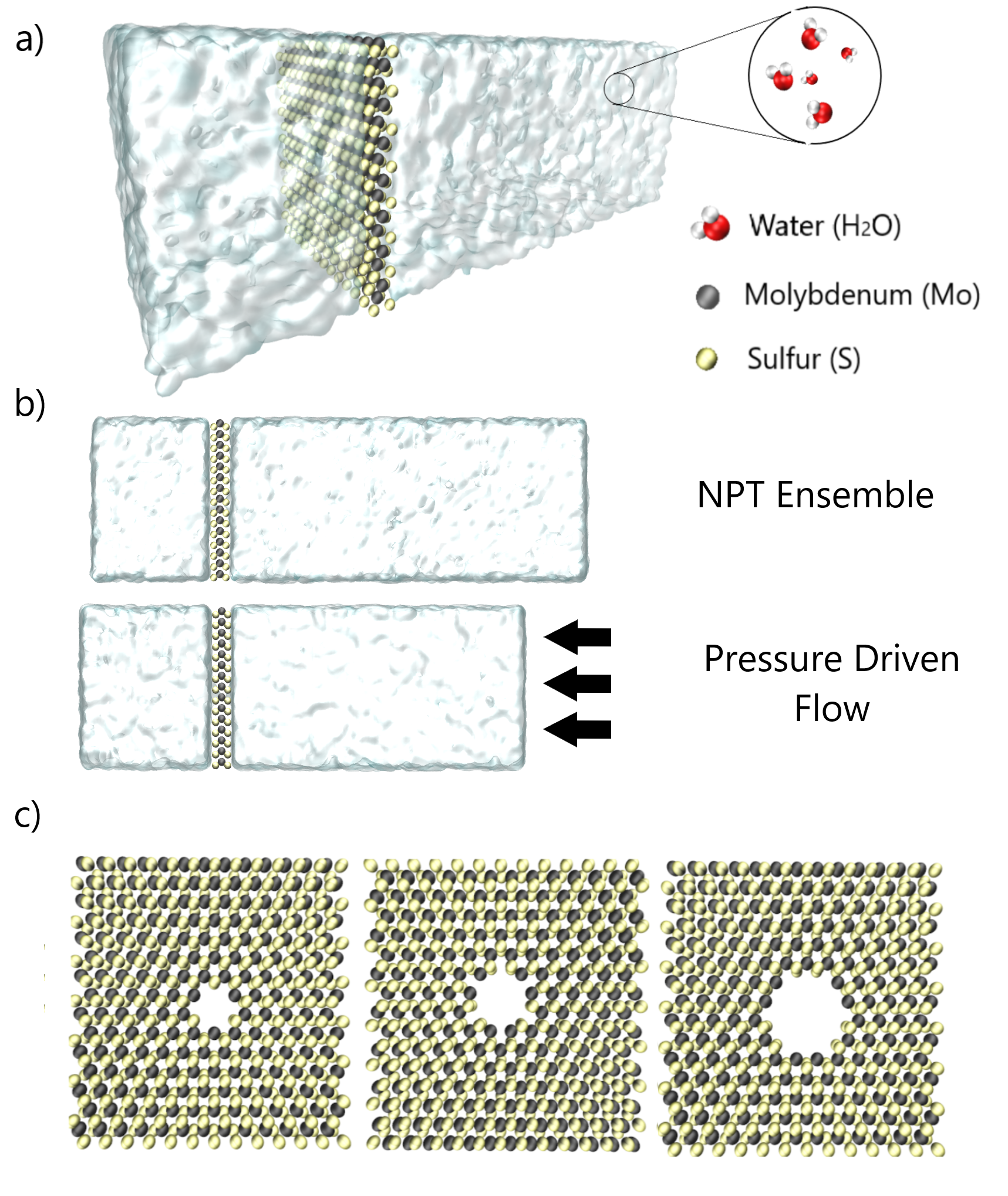} \caption{\textbf{(a)} Simulation box with the MoS$_2$ nanoporous membrane, with water in both reservoirs and graphene layers as pistons applying pressure (image created using the VMD software~\cite{HUMP96}). \textbf{(b)} The illustration of the simulation steps: the NPT equilibration and the following pressure driven process. \textbf{(c)} The 0.74nm, 0.97nm, and 1.33nm nanopore diameters from left to right (considering the center-to-center distance of atoms).} \label{fig1} \end{figure}

\begin{figure*}[hbt!] \centering
\includegraphics[width=16cm]{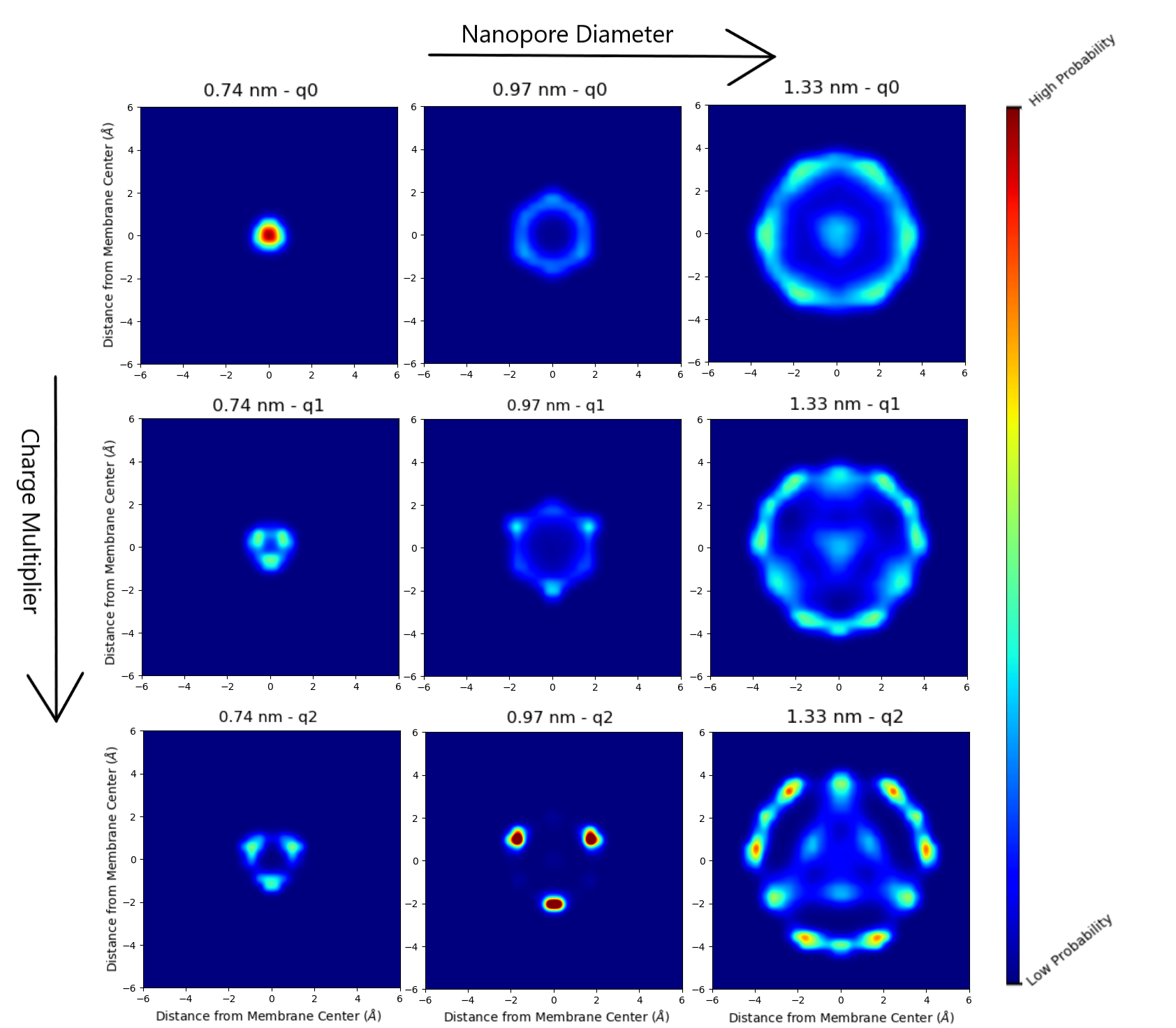} \caption{ Water flow density map (oxygen shown) for three different nanopore diameters and three different nanopore charge values.} \label{fig2} 
\end{figure*} 

The observations that the flux through membranes scales inversely with membrane thickness led to the idea of employing monolayer membranes as a new strategy for desalination. Posteriorly, several emerging classes of single-layer membranes have been proposed. Initially, exfoliated graphene as a single atomic layer membrane was proposed, followed by functionalized nanoporous graphene sheets~\cite{cohen-tanugi2012,konatham2013} with several active groups and inorganic nanoparticles. The first attempt to investigate functionalized graphene for desalination demonstrated that functionalized nanoporous graphene membranes could perform more than 99$\%$ salt rejection and provide water permeance up to 2 or 3 orders of magnitude higher than that of current commercially available reverse osmosis membranes and nanofiltration membranes~\cite{cohen-tanugi2012,wang2017}. Since in the case of CNTs the frictionless flow in 2nm diameter pores explain the high permeability of water and the hydrophilic functionalization, even though imposing additional friction helps rejecting salt, the mechanism and ingredients required for selectivity become clear. For atomic-sized pores, the concept of friction cannot be used. The selectivity observed in this case occurs for pores with one order of magnitude smaller than CNT pores. Therefore, although single-layer membranes seem to be the future for desalination, the physical-chemical reason for this enhancement flow is not clear.

\begin{figure*}[tbp!] \centering
\includegraphics[width=16cm]{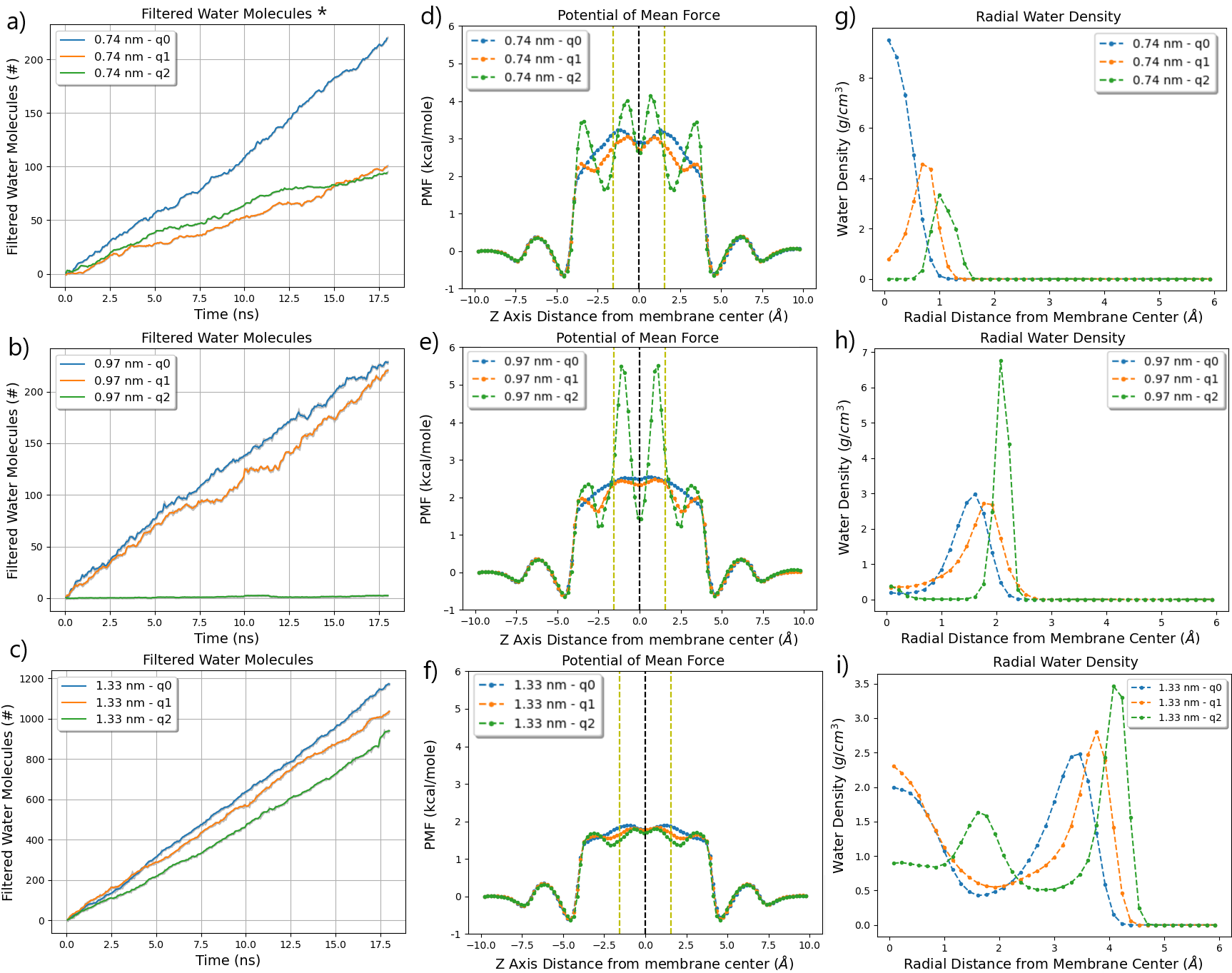} \caption{\textbf{(a-c)} Filtered water molecules in function of time, (*) refers to a simulation with pressure gradient of 2000 bar for statistical purposes,\textbf{(d-f)} the potential of mean force (PMF) of water near and inside the nanopore (vertical dashed lines represent the Mo-black and S-yellow positions) and \textbf{(g-h)} radial water density inside the nanopore for each combination of size and charge values.} \label{fig3} 
\end{figure*} 

The physics behind fast water movement inside a single-layer pore might involve a number of effects usually irrelevant in larger systems: the dimensions inside the pore are comparable with the screening Debye length, which enhances electrostatic effects, and is also comparable with the unrestricted path between molecular collisions in water and with pore surface anisotropies~\cite{Wang20177}. In this context, many emerging classes of single-layer membranes have also been advanced, beginning with functionalized nanoporous graphene sheets~\cite{cohen-tanugi2012,konatham2013,wang2017}. Contrary to what happens in CNTs, water permeability for equal pore diameter is larger for hydrophilic groups than when compared with hydrophobic ones~\cite{cohen-tanugi2012,konatham2013,wang2017,jafarzadeh2020}. 

Water, a polar molecule, is naturally attracted to the charged hydrophilic pore surface. However, what is puzzling about this process is why the water molecules, once connected with the hydrophilic wall surface, move away from it in a higher velocity than that observed in a hydrophobic pore with the same pressure. The  electrostatics of the system seem to play two competing roles: one to attract water to the pore, and another to repeal. These competing factors are enhanced in the case of the MoS$_2$ membrane~\cite{heiranian2015,Farimani-etal,liu-etal}. This system has a higher water permeability than its equivalent graphene~\cite{Farimani-etal}, probably because the charges distributed in the MoS$_2$ membrane attract water to the vicinity of the pore. The surprising effect is that this MoS$_2$ pore permeability is larger if the exposed atom at the pore is the divalent Mo when compared with the exposing monovalent S. Since pore sizes and shapes in both cases (Mo or S exposed) are slightly different, it is unclear if a larger exposed charge would enhance water velocity inside the pore, or if this larger velocity would be solely the result of its size and shape.

\begin{figure*}[hbt!] \centering
\includegraphics[width=16cm]{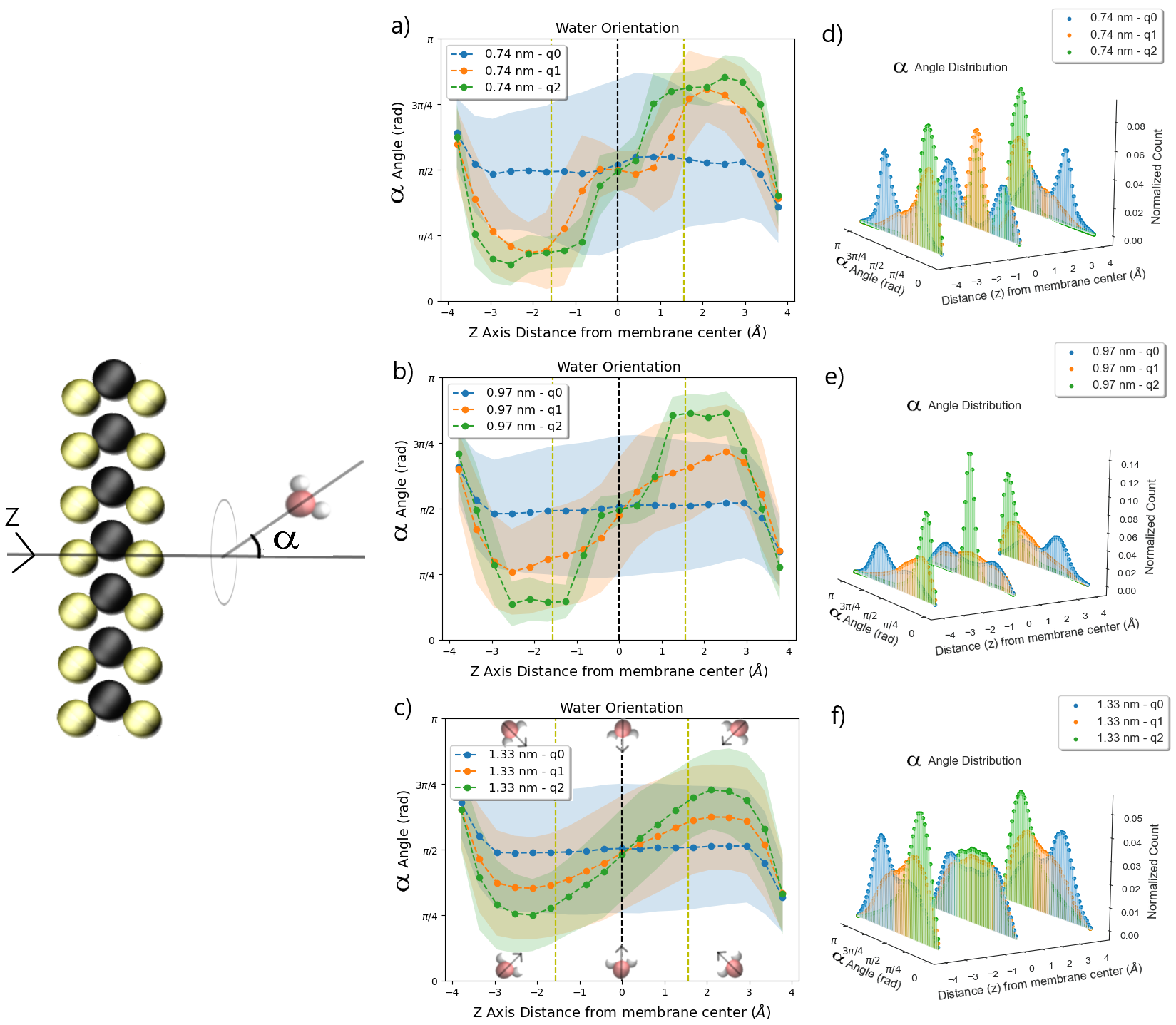} \caption{\textbf{(a-c)} The water molecule orientation respect to the z axis as function of the membrane center distance and \textbf{(d-f)} the $\alpha$ angle distribution inside the nanopore.} \label{fig4}
\end{figure*}

In order to clarify these questions, we performed nonequilibrium molecular dynamic (NEMD) simulations for an MoS$_2$ membrane. In earlier attempts to answer to this question, the comparison was between graphene pores with charge-altering functionalization. The issue with this procedure was that, in order to alter the charge, the functionalization also changed the available pore area and shape. We created a simulation box full of interagent particles, as described in Figure~\ref{fig1}(a), where two reservoirs of water molecules were separated by a MoS$_2$ nanoporous membrane. Through applying a pressure gradient in the box, the pressure-driven transport along the membrane initiates, as illustrated in Figure~\ref{fig1}(b). This type of process helps us to get insights toward designing new membrane materials and to better understand the water-nanopore relationship. 

To simply illustrate the complexity of tracking this problem, here we present the water permeability of an MoS$_2$ membrane under three conditions: a pore with no charge distribution (named $q_0$), a pore with Mo and S atoms with its normal charges (named $q_1$), and a pore with double the charge of Mo and S (named $q_2$). The first and last configurations are not physical, but they help us to understand the role played by the charges in water mobility without changing pore size and shape, while also keeping intact the spacial distribution of the charge in the membrane. 

In addition, three nanopore sizes were studied: 0.74nm, 0.97nm, and 1.33nm nanopore diameters (considering the center-to-center distance of atoms), as described in Figure~\ref{fig1}(c). The NEMD simulations were performed using the LAMMPS package~\cite{PLIMPTON19951}. We utilized the TIP4P/$\epsilon$~\cite{FUENTESAZCATL201686} water model and the parametrization of a reactive many-body potential as standard LJ parameters and charges values for Mo and S~\cite{KADANTSEV2012909}. The Lorentz-Berthelot mixing rules were used for non-bonded interactions. Long-range electrostatic interactions were calculated by the particle-particle-particle mesh method. We created a simulation box with 4nm x 4nm x 7nm, as illustrated in Figure~\ref{fig2}, with 3000 rigid water molecules. First, the particles were balanced during 2ns in an NPT ensemble in order to reach its $\approx$~1~$g/{cm}^3$ equilibrium density at 300K and 1atm. Graphene pistons were used to control applied pressures. Afterwards, a 500 bar of pressure was imposed in the feed reservoir and the 18ns of non-equilibrium molecular dynamics (NEMD) running starts. Transported water molecules were collected on the other side of the membrane. The results were averaged over 3 different set of simulations.

The water flowrate along MoS$_2$ illustrated in Figures~\ref{fig3}(a)-(c) is affected by two competing factors: charge and size. In general, the uncharged system presents the fastest flow. The charged pores, $q_1$ and $q_2$, form water clusters around Mo and S, as the shown in the color maps in Figure~\ref{fig2}. This behavior is similar to the observed functionalized carbon nanotubes~\cite{C8TA10941A}, graphene nanopores~\cite{cohen-tanugi2012,wang2017} and AlPO$_4$-54 nanotubes~\cite{gavazzoni2017}, and it is dominated by electrostatics. However, for the 0.97nm pore something unexpected happens. The $q_1$ charge pore has practically the same mobility as the uncharged system. At this size, the electrostatic interaction attracting water to Mo and S is probably overcome by the hydrogen-bond network, which forces water to move apart from Mo and S. At this same size, but with $q_2$ charges, water gets stuck inside the nanopore. 

The potential of mean force (PMF) illustrated in Figures~\ref{fig3}(d)-(f) shows that for the $q_2$ case PMF has minima in Mo and near S sites. A small charge decrease from $q_2$ to $q_1$ flattens the PMF, which can only be understood if another force overcomes the electrostatic interactions and if water molecules move away from Mo and S as shown in the Figures~\ref{fig3}(g)-(i). Also, from the PMFs shown in Figures~\ref{fig3}(d)-(f) we can see the implications of tunning the nanopore charge are local with short range.

To clarify what happens with the water flux in the 0.74nm and 0.97nm pore, we conducted a series of angular analysis to better understand the entrance and exit effects of enhanced dipole interaction between water molecules and the tuned membrane. As we can see from Figures~\ref{fig3}(a), the presence of a charge distribution when the nanopore is too small, practically independent of its strength, is enough to slow down the water mobility but it is not enough to trap the water molecules there, in contrast with what happens in the Figures~\ref{fig3}(b). 

Figures~\ref{fig4}(a)-(b)  highlight the molecular rotation around and inside the nanopore. From that, it is clear the effects of $q_2$ and $q_1$ compared to the $q_0$ case: the charge distribution implies in dipole-dipole interaction which is responsible for limiting the possible angular configurations to pass through the pore. 

Figures~\ref{fig4}(d)-(e) represent the evolution of the $\alpha$ angle distribution along the z axis. As we can see from it, the $q_0$ shows a range of angular possibilities and the rotation is not mandatory to pass through the pore while the $q_2$ and $q_1$ cases forces the molecule to rotate $\approx$~$90^{\circ}$ to be able to travel through the membrane. The angular constraint surely impacts the water flow-rate. Sometimes, the constraint is so limited that the water molecules gets trapped inside the nanopore, as illustrated by Figures~\ref{fig2}, Figures~\ref{fig3}(b) and Figures~\ref{fig4}(b)-(e) for the 0.97nm pore. In this case, the nanopore is larger so a higher number of water molecules fits in there. Attached to it, the hydrogen bond network is enhanced by the strong dipole interaction and contribute to the structure.    

It is interesting to note from previous studies~\cite{konatham2013,cohen-tanugi2012,C8TA10941A} that the ion selectivity is improved when the nanopore is functionalized. The functionalization procedure adds a charge distribution to the system and it improves the ion rejection by the membrane. As we can see from Figures~\ref{fig3}(a)-(c), the addition of charges slow down the water flowrate in all nanopores sizes, but it is a well paid costs thinking in ion selectivity if the charge distribution is enough to allow practically the same water flux as observed by the comparison between $q_0$ and $q_1$ cases in Figures~\ref{fig3}(b)-(c). In contrast, the cost is too high if the charge distribution can induce the nanopore blocking by strong water dipole interaction, as illustrated in Figures~\ref{fig3}(b)-(e)-(h).

The mechanism for high water mobility inside nanopores involve the competition between local electrostatic forces and the cooperation of the hydrogen bond network. The strong dipole interaction is responsible to append a constraint in the the water angular possibilities to travel through the nanopore, but this implication also depends on the nanopore size. This is the new trade-off in thinking about design the next-generation of nanoporous membrane materials. These nanopores needs a charge distribution in order to enhance the ion rejection, but not too much charge to trap water molecules into there depending on the nanopore size. The pore size, charge and shape surely impacts membrane permeability performance by different flow mechanisms.

\section{Acknowledgement}

The authors thank the financial support from the Brazilian agencies CNPq (through INCT-Fcx), Coordena\c{c}\~ao de Aperfei\c{c}oamento de Pessoal de N\'ivel Superior (CAPES) and the computational infrastructure CENAPAD/SP and CESUP/UFRGS.

\bibliographystyle{unsrt}
\bibliography{aapmsamp}

\begin{thebibliography}{10}

\bibitem{VOUTCHKOV20182}
Nikolay Voutchkov.
\newblock Energy use for membrane seawater desalination – current status and
  trends.
\newblock {\em Desalination}, 431:2 -- 14, 2018.

\bibitem{Alvarez2018}
Pedro J.~J. Alvarez, Candace~K. Chan, Menachem Elimelech, Naomi~J. Halas, and
  Dino Villagr{\'a}n.
\newblock Emerging opportunities for nanotechnology to enhance water security.
\newblock {\em Nature Nanotechnology}, 13(8):634--641, Aug 2018.

\bibitem{Werber2016}
Jay~R. Werber, Chinedum~O. Osuji, and Menachem Elimelech.
\newblock Materials for next-generation desalination and water purification
  membranes.
\newblock {\em Nature Reviews Materials}, 1(5):16018, Apr 2016.

\bibitem{hummer2001}
G.~Hummer, J.~C Rasiah, and J.~P. Nowryta.
\newblock Water conduction through the hydrophobic channel of a carbon
  nanotube.
\newblock {\em Nature}, 414:188--190, 2001.

\bibitem{majumder2005}
M.~Majumder, N.~Chopra, R.~Andrews, and B.~J. Hinds.
\newblock Enhanced flow in carbon nanotubes.
\newblock {\em Nature}, 438:44, 2005.

\bibitem{holt2006}
J.~K. Holt, H.~G. Park, Y.~Wang, M.~Stadermann, A.~B. Artyukhin, C.~P
  Grigoropoulos, Noy A., and O.~Bakajin.
\newblock Fast mass transport through sub-2-nanometer carbon nanotubes.
\newblock {\em Science}, 312:1034--1037, 2006.

\bibitem{kalra2002}
A.~Kalra, S.~Garde, and G.~Hummer.
\newblock Osmotic water transport through carbon nanotube membranes.
\newblock {\em Proc. of National Acad. Sciences}, 100:210175--10180, 2002.

\bibitem{song2009}
C.~Song and B.~Corry.
\newblock Intrinsic ion selectivity of narrow hydrophobic pores.
\newblock {\em J. Phys. Chem. B}, 113:7662--7649, 2009.

\bibitem{C8TA10941A}
Yang Hong, Jingchao Zhang, Chongqin Zhu, Xiao~Cheng Zeng, and Joseph~S.
  Francisco.
\newblock Water desalination through rim functionalized carbon nanotubes.
\newblock {\em J. Mater. Chem. A}, 7:3583--3591, 2019.

\bibitem{gallo2017}
M.~De~Marzio, G.~Camisasca, M.~M. Conde, M.~Rovere, and P.~Gallo.
\newblock Structural properties and fragile to strong transition in confined
  water.
\newblock {\em The Journal of Chemical Physics}, 146(8):084505, 2017.

\bibitem{HUMP96}
William Humphrey, Andrew Dalke, and Klaus Schulten.
\newblock {VMD} -- {V}isual {M}olecular {D}ynamics.
\newblock {\em Journal of Molecular Graphics}, 14:33--38, 1996.

\bibitem{cohen-tanugi2012}
David Cohen-Tanugi and Jeffrey~C. Grossman.
\newblock Water desalination across nanoporous graphene.
\newblock {\em Nano Lett.}, 12:3602--3608, 2012.

\bibitem{konatham2013}
D.~Konatham, J.~Yu, ?T.~A. Ho, and A.~Striolo.
\newblock Simulation insights for graphene-based water desalination membranes.
\newblock {\em Lamgmuir}, 29:11884--11897, 2013.

\bibitem{wang2017}
Y.~Wang, Z.~He, K.~M. Gupta, Q.~Shi, and Ruifeng Lu.
\newblock Molecular dynamics study on water desalination through functionalized
  nanoporous graphene.
\newblock {\em Carbon}, 12:120--127, 2017.

\bibitem{Wang20177}
Luda Wang, Michael S.~H. Boutilier, Piran~R. Kidambi, Doojoon Jang, Nicolas~G.
  Hadjiconstantinou, and Rohit Karnik.
\newblock Fundamental transport mechanisms, fabrication and potential
  applications of nanoporous atomically thin membranes.
\newblock {\em Nature Nanotechnology}, 12(6):509--522, Jun 2017.

\bibitem{jafarzadeh2020}
R~Jafarzadeh, J.~Azamat, and H.~Erfan-Niya.
\newblock Water desalination across functionalized silicon carbide nanosheet
  membranes: insights from molecular simulations.
\newblock {\em Estrutural Chemistry}, 31:293--303, 2020.

\bibitem{heiranian2015}
M.~Heiranina, A.~B. Farimani, and N.~R.. Aluru.
\newblock Water desalination with a single-layer mos$_2$ nanopore.
\newblock {\em Nature Communications}, 6:8616--8622, 2015.

\bibitem{Farimani-etal}
Zhonglin Cao, Vincent Liu, and Amir Barati~Farimani.
\newblock Why is single-layer mos2 a more energy efficient membrane for water
  desalination?
\newblock {\em ACS Energy Letters}, 5(7):2217--2222, 2020.

\bibitem{liu-etal}
Chunjiao Liu, Yakang Jin, and Zhigang Li.
\newblock Water transport through graphene and mos2 nanopores.
\newblock {\em Journal of Applied Physics}, 126(2):024901, 2019.

\bibitem{PLIMPTON19951}
Steve Plimpton.
\newblock Fast parallel algorithms for short-range molecular dynamics.
\newblock {\em Journal of Computational Physics}, 117(1):1 -- 19, 1995.

\bibitem{FUENTESAZCATL201686}
Raúl Fuentes-Azcatl and Marcia~C. Barbosa.
\newblock Thermodynamic and dynamic anomalous behavior in the tip4p/e water
  model.
\newblock {\em Physica A: Statistical Mechanics and its Applications}, 444:86
  -- 94, 2016.

\bibitem{KADANTSEV2012909}
Eugene~S. Kadantsev and Pawel Hawrylak.
\newblock Electronic structure of a single mos2 monolayer.
\newblock {\em Solid State Communications}, 152(10):909 -- 913, 2012.

\bibitem{gavazzoni2017}
Cristina Gavazzoni, Nicolas Giovambattista, Paulo~A. Netz, and Marcia~C.
  Barbosa.
\newblock Structure and mobility of water confined in alpo4-54 nanotubes.
\newblock {\em The Journal of Chemical Physics}, 146(23):234509, 2017.

\end{thebibliography}

\end{document}